\documentclass[10pt]{book}  

\usepackage{sussp2004}
\usepackage{graphicx}
\usepackage{latexsym}  
\usepackage{makeidx}  

\usepackage[english]{babel}  

\newcommand{\half}{\mbox{\small $\frac{1}{2}$}}

\begin{document}  

\setcounter{section}{1}
\setcounter{page}{1}

\headings{Chiral extrapolations}{Chiral extrapolations}
{Meinulf G\"ockeler}
{Institut f\"ur Theoretische Physik \\ Universit\"at Leipzig}

\section{Introduction}

\vspace*{-10.2cm}
LU-ITP 2004/047 
\vspace*{9.8cm}

\noindent
The title `Chiral extrapolations'\index{chiral extrapolation} of this 
contribution refers 
more precisely to chiral extrapolations {\emph {of lattice {\sc{qcd}} data}}. 
We shall deal with low-energy aspects of {\sc{qcd}}, which are not accessible
to ordinary weak-coupling perturbation theory. Possible alternatives to 
weak-coupling perturbation theory in the low-energy domain of {\sc{qcd}} 
include the investigation of specific models, Monte Carlo simulations 
of lattice regularised {\sc{qcd}}, and chiral effective field 
theories\index{chiral effectice field theory} 
({\sc{cheft}}), the latter being low-energy theories which incorporate 
the constraints from (spontaneously broken) chiral 
symmetry\index{chiral symmetry}. It is the 
comparison of {\sc{cheft}} (or chiral perturbation 
theory\index{chiral perturbation theory} ({\sc{chpt}})) 
with results from lattice {\sc{qcd}} simulations that will be the 
subject of the present paper.

However, the reader should be warned that as lattice {\sc{qcd}} practitioners
we look at {\sc{cheft}} from an (ab)user's viewpoint. Also a second warning
may be in order: This is not a review. The material has been selected
according to subjective criteria, and the references are far from 
being complete. For a review of {\sc{chpt}} see, {\it e.g.}, 
(Leutwyler 2000) and (Mei{\char"FF}ner 2000). For reviews on closely related 
subjects, partly overlapping with the present work, see (B\"ar 2004) 
and (Colangelo 2004).

\section{Lattice regularisation and Monte Carlo 
         simulation\index{Monte Carlo simulation}}

The basic input for a Monte Carlo simulation of 
lattice {\sc{qcd}}\index{lattice QCD} is first
of all a lattice action, \ie a discretised version of Euclidean {\sc{qcd}}. 
Secondly, one has to choose a lattice size (necessarily finite). 
Thirdly, the bare 
coupling constant and the quark mass(es) have to be fixed. 
Then the lattice spacing $a$ and the spatial box size
$L$ can (approximately) be given in physical units.

In general it is not possible to choose the simulation parameters
such that the results can immediately be identified with 
experimentally measurable quantities. In particular, three extrapolations
are required: the continuum limit $a \to 0$, the thermodynamic limit
$L \to \infty$, and the chiral limit\index{chiral limit},
where the masses of the light quarks
decrease to their physical values and further down to zero. 
Unfortunately, in all three limits the simulation costs increase 
rapidly. It is therefore preferable to appeal to theory in order to relate
simulation results obtained for unphysical quark masses, finite 
volumes, {\ldots} to phenomenology. This will then lead to well-justified
extrapolation formulae.

\section{Chiral effective field theory in the pion 
         sector\index{chiral effective field theory}}

The natural starting point for chiral perturbation theory is the pion 
sector. The very existence of light pions (for the case of two flavours)
relies on the spontaneous breakdown of chiral symmetry combined with the
weak explicit breaking due to the quark masses. A further consequence is
the weakness of the pion-pion interaction at low energies and momenta
which makes a perturbative treatment meaningful. This (chiral) 
perturbation theory is most conveniently set up by means of an effective
Lagrangian, \ie the most general Lagrangian for effective pion (and later on
also nucleon, \ldots) fields which is compatible with chiral symmetry. 
The effective Lagrangian is constructed out of terms with more and more 
derivatives as the order of the expansion increases, leading at the same 
time to an increasing number of effective coupling constants, which are 
not determined by chiral symmetry. Eventually one obtains an expansion
of the physical observables in the pion mass $M_\pi$ and the particle 
momenta, where both are considered as quantities of the order of a
small parameter $p$. Here `small' means 
$p \ll 4 \pi F_\pi \sim 1 \, \mbox{GeV}$.

We shall restrict ourselves to the case of two flavours, $N_f = 2$, with
isospin breaking neglected, \ie we take for the quark masses 
$m_u = m_d = m_q$. Then the lowest-order expression for 
$M_\pi$ is the famous Gell-Mann--Oakes--Renner relation 
$M_\pi^2 = 2 |\langle \bar{q} q \rangle| m_q / F^2$, where 
$\langle \bar{q} q \rangle$ is the chiral condensate and $F$ denotes the 
pion decay constant $F_\pi$ in the chiral limit.

Beyond leading order one finds (see, {\it e.g.}, Colangelo \etal 2001)
\begin{equation}
M_\pi^2 = M^2 \left\{ 1 - \half x \hat{\ell}_3 
  + \mbox{\small $\frac{17}{8}$} x^2 \hat{\ell}_M^2 + x^2 k_M + O(x^3) 
     \right\} 
\label{eq:mpimq}
\end{equation}
with
\begin{equation}
\hat{\ell}_M = \frac{1}{51} (28 \hat{\ell}_1 + 32 \hat{\ell}_2 
              - 9 \hat{\ell}_3 + 49) \,.
\end{equation}
Here we have set $M^2 = 2 m_q B$ with $B = - \langle \bar{q} q \rangle / F^2$
and $x = M^2/(16 \pi^2 F^2)$. The chiral 
logarithms\index{chiral logarithms} are hidden in 
the quantities $\hat{\ell}_i = \ln (\Lambda_i^2/M^2)$, which contain the 
information on the (renormalised) coupling constants. The term proportional to
$k_M$ represents analytic contributions $O(x^2)$, which are expected to 
be small. Note that $k_M$ and $\Lambda_i$ depend neither on $m_q$ nor on 
the renormalisation scale. One can estimate $F = 0.0862 \, \mbox{GeV}$, 
and phenomenological analyses lead to 
$\Lambda_1 = 0.12_{-0.03}^{+0.04} \, \mbox{GeV}$,
$\Lambda_2 = 1.20_{-0.06}^{+0.06} \, \mbox{GeV}$,
$\Lambda_3 = 0.59_{-0.41}^{+1.40} \, \mbox{GeV}$,
$\Lambda_4 = 1.25_{-0.13}^{+0.15} \, \mbox{GeV}$ 
(see, {\it e.g.}, Colangelo and D\"urr (2004)).

\section{Comparison with Monte Carlo data in the pion sector}

Let us start with some general remarks on the comparison of Monte Carlo
data with {\sc{cheft}} formulae. First of all, one needs results in physical
units. A popular way to set the physical scale\index{physical scale} 
uses the Sommer parameter
$r_0$, which is a length scale derived from the heavy-quark potential $V(R)$
through the condition $\mathrm d V(R) / \mathrm d R |_{R=r_0} = 1.65 $.
The phenomenological value has been found to be approximately 
$r_0 = 0.5 \, \mbox{fm}$, which is the number to be used in the following.
Note that this method assumes that the dependence of $r_0$ on the light 
quark masses is negligible, an assumption whose validity is not quite clear.
Secondly, many quantities, such as, {\it e.g.}, quark masses, have to be 
renormalised.
Finally, we have to deal with lattice artefacts. Ideally, 
one would eliminate them by an extrapolation to the continuum limit, which
is not an easy task, or one could incorporate them in the {\sc{cheft}} (see the
review by B\"ar (2004)). In the following we shall adopt a simple-minded 
approach and try to select the data such that cut-off effects are negligible. 

Typically there is little structure in the 
quark-mass dependence of lattice results, and the data are in most cases 
well described by a linear function of $m_q$. In other words,
there are no obvious chiral logarithms\index{chiral logarithms}. 
Of course, this may be due to the
relatively large quark masses in the present simulations, where leading
order {\sc{chpt}} is unlikely to work. Thus one needs higher-order 
calculations, and one has to face the question up to which masses 
{\sc{chpt}} is reliable. Alternatively, one may
try to tame the unphysical behaviour of the truncated series 
at large masses by some cut-off function and in this
way arrive at a formula which works in the mass range covered by the 
simulations. For this approach see, {\it e.g.}, (Leinweber \etal 2004), 
(Young \etal 2004), and references therein.

\begin{figure}
\centering{
\includegraphics[width=11.4cm,clip]{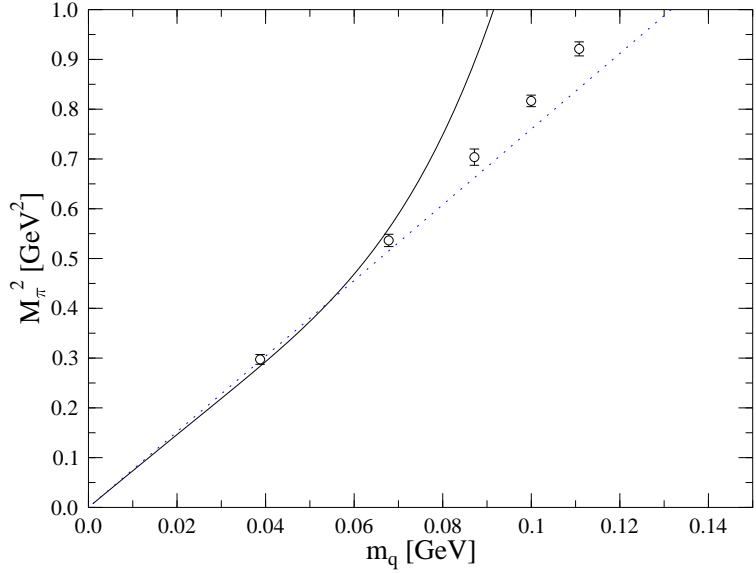}
\caption{The square of the pion mass versus the quark mass. {\sc{jlqcd}} data
         for $N_f=2$ are compared with the Gell-Mann--Oakes--Renner relation 
         (dotted line) and chiral perturbation theory (full line).}
\label{fig:mpimq}
}
\end{figure}

In Fig.~\ref{fig:mpimq} we compare pion and quark 
masses\index{pion mass} obtained by the
{\sc{jlqcd}} collaboration (Aoki \etal 2003) with the quark-mass dependence
of the pion mass as predicted by Eq.~(\ref{eq:mpimq}). The parameters have
been chosen (not fitted) as follows: $F = 0.0862 \, \mbox{GeV}$, 
$B = 3.8 \, \mbox{GeV}$,
$\Lambda_1 = 0.12 \, \mbox{GeV}$, $\Lambda_2 = 1.20 \, \mbox{GeV}$,
$\Lambda_3 = 0.65 \, \mbox{GeV}$, $\Lambda_4 = 1.25 \, \mbox{GeV}$.
It is a remarkable observation that the 
Gell-Mann--Oakes--Renner relation\index{Gell-Mann--Oakes--Renner relation}
$M_\pi^2 = 2 B m_q$ is a 
rather good approximation for pion masses up to about $0.7 \, \mbox{GeV}$.
The above parameters are well compatible with the phenomenological
values quoted in the preceding section. Only $B$ is somewhat larger than
the value $B = 2.8 \, \mbox{GeV}$ given in (D\"urr 2003). Note that $B$
is scale and scheme dependent as are the quark masses, while the product
$B m_q$ is independent of scale and scheme. Here we have employed tadpole 
improved one-loop perturbation theory with the renormalisation scale
$\mu = 2 \, \mbox{GeV}$ in order to convert the bare VWI masses to 
renormalised quark masses $m_q$ in the $\overline{\mbox{MS}}$ scheme. 
The use of perturbation theory entails a considerable uncertainty in 
the renormalised quark masses and hence in $B$. Indeed, a recent
investigation (G\"ockeler \etal 2004) suggests that the non-perturbative
mass renormalisation factor (at the bare coupling used by the {\sc{jlqcd}} 
collaboration) is about 2.3 times larger than the perturbative estimate 
employed here. But it is gratifying to
see that chiral perturbation theory with phenomenologically acceptable
values of the coupling constants is able to make contact with the 
low-mass end of the quark mass range that can be reached in present 
simulations with dynamical quarks. For a more detailed discussion of 
the quark-mass dependence of $M_\pi$ in comparison with different 
Monte Carlo data see (D\"urr 2003).

\begin{figure}
\centering{
\includegraphics[width=11.4cm,clip]{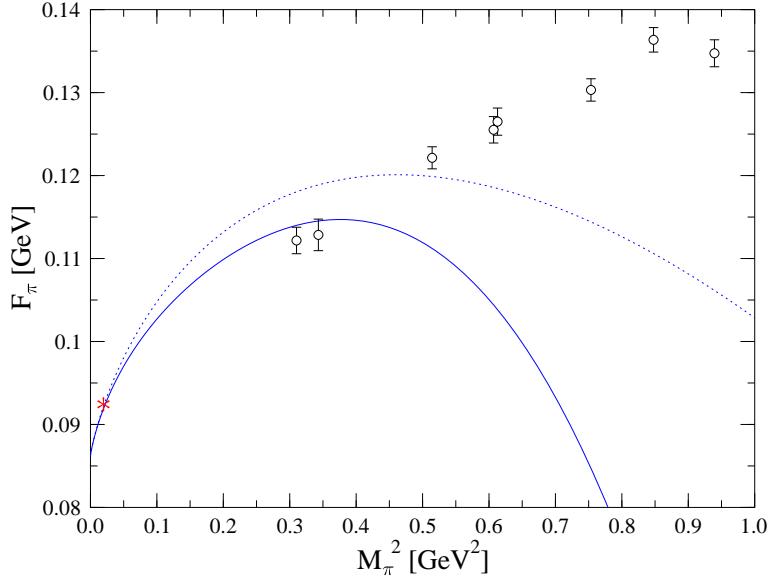}
\caption{Preliminary data ($N_f=2$) for the pion decay constant compared with
         chiral perturbation theory at different orders. The asterisk
         indicates the physical point.}
\label{fig:fpi}
}
\end{figure}

For the pion decay constant\index{pion decay constant} 
$F_\pi$ (normalised such that the physical 
value is $92.4 \, \mbox{MeV}$) chiral perturbation theory yields
(see, {\it e.g.}, Colangelo \etal 2001)
\begin{equation} 
F_\pi = F \left\{ 1 + x \hat{\ell}_4 
  - \mbox{\small $\frac{5}{4}$} x^2 \hat{\ell}_F^2 + x^2 k_F + O(x^3) \right\}
\label{eq:fpimq}
\end{equation} 
with 
\begin{equation} 
\hat{\ell}_F = \frac{1}{30} (14 \hat{\ell}_1 + 16 \hat{\ell}_2 
              + 6 \hat{\ell}_3 - 6 \hat{\ell}_4 + 23) 
             = \ln \frac{\Lambda_F^2}{M^2} \,.
\end{equation} 
Again, the analytic contributions $O(x^2)$ (proportional to $k_F$) are
expected to be small. In Fig.~\ref{fig:fpi} we compare this formula with
preliminary data from the {\sc{ukqcd}} and {\sc{qcdsf}} collaborations. 
Motivated by our observation that the Gell-Mann--Oakes--Renner relation 
works so well, we replace $M$ by $M_\pi$. Thus we avoid the problem 
of renormalising 
the quark mass. Choosing $F = 0.0862 \, \mbox{GeV}$, $k_F = -0.5$, 
$\Lambda_4 = 1.12 \, \mbox{GeV}$, $\Lambda_F = 0.67 \, \mbox{GeV}$
(consistent with phenomenology) we obtain the full curve in 
Fig.~\ref{fig:fpi}, which connects the physical point with the data
for the lowest mass but does not describe the data at larger masses. 
It is however reassuring that for masses up to the first Monte Carlo 
points the chiral expansion seems to be well-convergent: The dotted curve,
which corresponds to $F_\pi = F \{ 1 + x \hat{\ell}_4 \}$, does not 
deviate dramatically from the full curve in this region.

\section{Including baryons}
\label{sec:bary}

The nucleon mass $m_N$ does not vanish even in the chiral limit. Indeed
$m_N \gg M_\pi$, and a non-relativistic treatment of the nucleon field 
seems reasonable. This leads to the so-called heavy-baryon chiral 
perturbation theory\index{heavy-baryon chiral perturbation theory}
({\sc{hbchpt}}). A relativistic formulation of chiral 
perturbation theory in the baryon sector has been given
by Becher and Leutwyler (1999). In both cases the physical picture is that
of a (heavy) nucleon core surrounded by a cloud of light pions. 
This is rather different from the situation for the pion, and 
hence the behaviour of the chiral expansion in the nucleon sector
need not be similar to that in the pion sector.

In Becher's and Leutwyler's formulation one obtains for the nucleon 
mass\index{nucleon mass} (Becher and Leutwyler 1999, Procura \etal 2004)
\begin{equation} \begin{array}{l} \displaystyle
 m_N  =  m_0 - 4 c_1 M_\pi^2-\frac{3 (g_A^0)^2}{32 \pi F^2} M_\pi^3 
 + \left[e_1^r(\lambda)-\frac{3}{64 \pi^2 F^2}
    \left( \frac{(g_A^0)^2}{m_0} - \frac{c_2}{2} \right) \right.
\\[0.8cm] \displaystyle \hspace*{0.1cm}  \left. {}
  - \frac{3}{32 \pi^2 F^2}
       \left( \frac{(g_A^0)^2}{m_0} - 8c_1 + c_2 + 4 c_3 \right)
   \ln{\frac{M_\pi}{\lambda}} \right] M_\pi^4
 + \frac{3 (g_A^0)^2}{256 \pi F^2 m_0^2}M_\pi^5 + O(M_\pi^6) \,.
\end{array}
\label{eq:mnfit}
\end{equation}
Here we have again identified $M_\pi^2 = M^2 = 2 B m_q$, $\lambda$ is the  
renormalisation scale, $m_0$ and $g_A^0$ are the mass and the axial charge
of the nucleon in the chiral limit, $c_1$, $c_2$, $c_3$ denote coupling
constants from the effective Lagrangian, and $e_1^r(\lambda)$ is a 
counterterm. 

\begin{figure}
\centering{
\includegraphics[width=11.4cm,clip]{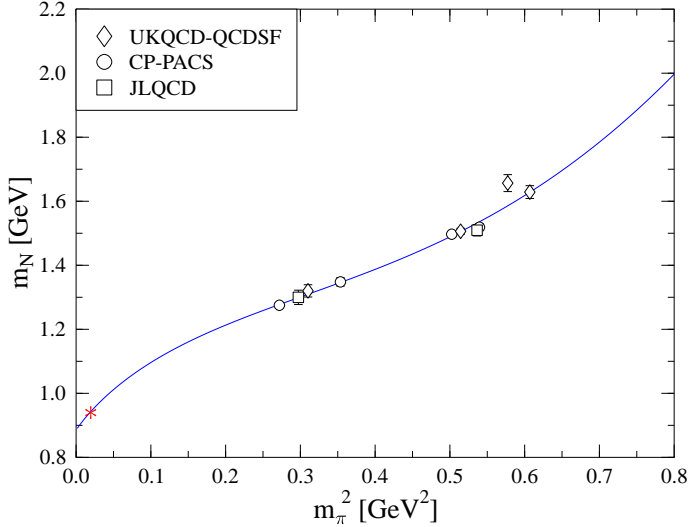}
\caption{Nucleon mass data for $N_f=2$ on (relatively) large and fine 
         lattices. The asterisk indicates the physical point. The curve 
         corresponds to the fit mentioned in the text.}
\label{fig:mnfit}
}
\end{figure}

Hadron masses for $N_f=2$ have been published by the {\sc{cp-pacs}} 
collaboration (Ali~Khan \etal 2002), the {\sc{jlqcd}} collaboration 
(Aoki \etal 2003) as well as the {\sc{ukqcd}} and {\sc{qcdsf}} 
collaborations (Allton \etal 2002, Ali~Khan \etal 2004a).
We have selected results obtained on (relatively) large and fine lattices
to compare with chiral perturbation theory. More precisely, we have 
considered masses from simulations with $a < 0.15 \, \mbox{fm}$, 
$M_\pi < 800 \, \mbox{MeV}$ and $M_\pi L > 5$. These ten data points
were fitted with Eq.~(\ref{eq:mnfit}) where $g_A^0$, $F$, $c_2$ and $c_3$
were fixed to phenomenologically reasonable values while $m_0$, $c_1$
and $e_1^r(\lambda = 1 \, \mbox{GeV})$ were the fit parameters. For more 
details see (Ali~Khan \etal 2004a). The fit curve and the data points 
are shown in Fig.~\ref{fig:mnfit}. It is first of all remarkable that
the masses obtained by different collaborations with different lattice 
actions and algorithms fall (to rather good accuracy) onto a single curve. 
Furthermore, the fit parameters are very well compatible with phenomenology,
in particular, $c_1$ is about $-1 \, \mbox{GeV}^{-1}$ and the fit curve 
comes quite close to the physical point. On
the other hand, Eq.~(\ref{eq:mnfit}) seems to work up to surprisingly
large masses.

\section{Axial charge of the nucleon\index{axial charge}}

The quark-mass dependence of the axial charge (or axial-vector coupling
constant\index{axial-vector coupling constant}) $g_A$ of the 
nucleon has been studied by 
Hemmert \etal (2003) within the framework of the so-called small-scale
expansion\index{small-scale expansion}. This is an extension 
of {\sc{hbchpt}} which includes explicit 
$\Delta(1232)$ degrees of freedom. 

In the small-scale expansion the expansion parameter is usually called
$\epsilon$, and one finds in $O(\epsilon^3)$:
\begin{equation} \begin{array}{l} \displaystyle
g_A (M_\pi^2) = 
  g^0_A-\frac{(g^0_A)^3M_\pi^2}{16\pi^2 F^2}
\\ \displaystyle \quad {}
  +4\left\{C^{SSE}(\lambda)
  +\frac{c_A^2}{4\pi^2 F^2}\left[\frac{155}{972}\, g_1
  -\frac{17}{36}\, g^0_A\right]+\gamma^{SSE}
   \ln{\frac{M_\pi}{\lambda }}\right\}M_\pi^2
\\ \displaystyle \quad {}
   +\frac{4c_A^2 
   g^0_A}{27 \pi F^2 \Delta}M_\pi^3
   +\frac{8}{27\pi^2 F^2}\;c_A^2 g^0_A M_\pi^2
   \sqrt{1-\frac{M_\pi^2}{\Delta^2}}\,\ln{R}
\\ \displaystyle \quad {}
   +\frac{c_A^2\Delta^2}{81\pi^2 F^2}\left(25g_1-
   57g_A^0\right)\left\{\ln\left[\frac{2\Delta}{M_\pi}
   \right]-\sqrt{1-\frac{M_\pi^2}{\Delta^2}}\ln R\right\}
   + O (\epsilon^4)
\end{array}
\label{eq:ga}
\end{equation}
with
\begin{equation} 
\gamma^{SSE} = \frac{1}{16\pi^2 F^2}
\left[\frac{50}{81}\,c_A^2 g_1-\frac{1}{2}\,g_A^0
   -\frac{2}{9}\,c_A^2g_A^0-(g_A^0)^3\right]
\,, \,
R = \frac{\Delta}{M_\pi }+\sqrt{\frac{\Delta^2}{M_\pi^2}-1} \,.
\end{equation}
The new parameters appearing here are $g_A^0$ (the value of $g_A$ in the
chiral limit), $\Delta$ (the nucleon $\Delta$ mass splitting in the 
chiral limit), $c_A$, $g_1$ (the $N \Delta$ and $\Delta \Delta$ axial 
coupling constants), and $C^{SSE}(\lambda)$ (a counterterm at the 
renormalisation scale $\lambda$). Hemmert \etal (2003) find that
for reasonable values of the parameters the formula (\ref{eq:ga}) is
able to describe the rather weak mass dependence of the Monte Carlo 
data as well as the physical point. 

\section{Chiral effective field theory in a finite 
         volume\index{chiral effective field theory}}

Presently, lattice {\sc{qcd}} simulations are not only restricted 
to unphysical quark masses, but also to relatively small (spatial) 
volumes, usually with periodic boundary conditions. While simulations 
at the physical quark
masses might be possible some day, it will take a bit longer before the
ideal case of an infinite volume simulation can be realised. In the
meantime we can take advantage of the fact that the chiral effective
Lagrangian is volume independent for periodic boundary conditions
(Gasser and Leutwyler 1988). So the 
same Lagrangian governs the quark-mass as well as the volume dependence, 
and additional information on the coupling constants can be extracted 
from finite size effects. 
This description of the finite size effects\index{finite size effects} 
should work as long as they
result from the deformation of the pion cloud 
in the finite volume, \ie as long as $L$ is not too small. After all,
it is the pion propagation that is predominantly affected by the finite 
volume, because the pion is the lightest particle in the theory.
Treating $L^{-1}$ as a quantity of order $p$ like 
$M_\pi$ we arrive at the so-called $p$ expansion
(Gasser and Leutwyler 1987). This is to
be distinguished from the $\epsilon$ expansion, where $L^{-1} = O(\epsilon)$,
$M_\pi = O(\epsilon^2)$ with a small parameter $\epsilon$. 

In more technical terms, the finite volume (with periodic boundary 
conditions) discretises the allowed momenta such that the momentum 
components are restricted to integer multiples of $2 \pi / L$. So the loop
integrals of {\sc{chpt}} become sums. On the other hand, we can interpret the
resulting expressions in the following way. 
In a finite volume a pion emitted from
a nucleon can not only be reabsorbed by the same nucleon, but also  
by one of the periodic images of the original nucleon at 
a distance which is an integer multiple of $L$ in each of the finite 
directions. From this point of view the finite size 
effects\index{finite size effects} arise from
pions travelling around the volume once, twice, {\ldots} in a given direction,
and each crossing of the boundary leads (roughly) to a factor 
$\exp (-M_\pi L)$ in the final contribution.

\begin{figure}
\centering{
\includegraphics[width=11.4cm,clip]{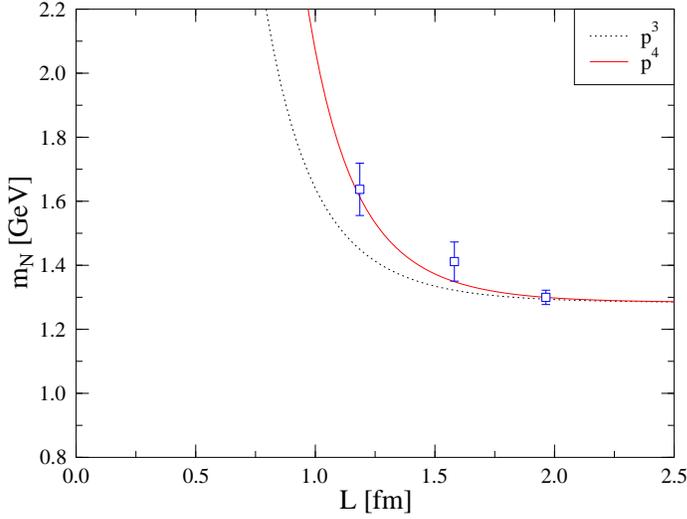}
\caption{Volume dependence of the nucleon mass for 
$M_\pi = 545 \, \mbox{MeV}$. The dotted curve shows the contribution of
the $p^3$ term, while the solid curve includes also the $p^4$ correction.}
\label{fig:mn545fv}
}
\end{figure}

As an example, let us consider the nucleon mass\index{nucleon mass} 
in relativistic baryon {\sc{chpt}}. Then the nucleon mass in a 
spatial box of length $L$, but for an infinite extent in time direction, 
can be written as 
\begin{equation}
m_N(L) = m_N(\infty) + \Delta_a (L) + \Delta_b (L) + O(p^5),
\end{equation}
where 
\begin{equation}
\Delta_a(L) = \frac{3 (g_A^0)^2 m_0 M_\pi^2}{16 \pi^2 F^2}
 \int_0^\infty \! \mathrm d x \, \sum_{\vec{n} \neq \vec{0}} \, 
  K_0 \left( L |\vec{n}| \sqrt{m_0^2 x^2 + M_\pi^2 (1-x)} \right) 
\end{equation}
is the $O(p^3)$ contribution to the finite size effect and 
\begin{equation}
\Delta_b(L) = 
 \frac{3 M_\pi^4}{4 \pi^2 F^2} \sum_{\vec{n} \neq \vec{0}} 
\left[ (2 c_1 - c_3) \frac{K_1(L |\vec{n}| M_\pi)}{L |\vec{n}| M_\pi}
+ c_2 \frac{K_2(L |\vec{n}| M_\pi)}{(L |\vec{n}| M_\pi)^2} \right] 
\end{equation}
is the additional contribution arising at $O(p^4)$. Here $n_i$ can be 
interpreted as the number of times the pion crosses the `boundary' 
of the box in the $i$ direction. Note that the finite volume does not
introduce any new coupling constant. 
For a comparison with
Monte Carlo data we evaluate the finite size corrections using the 
parameters from the fit discussed in Section~\ref{sec:bary}, choose 
$m_N(\infty)$ such that 
$m_N(L)$ on the largest lattice agrees with the Monte Carlo value and 
take for $M_\pi$ the value from the largest lattice. This leads to a
surprisingly good agreement with the Monte Carlo masses, as is shown
for $M_\pi = 545 \, \mbox{MeV}$ in Fig.~\ref{fig:mn545fv} using {\sc{jlqcd}} 
data. Even for larger pion masses the formula reproduces the finite
size effects quite well (Ali~Khan \etal 2004a). This lends further support
to the fit shown in Fig.~\ref{fig:mnfit}.

In the case of the pion mass\index{pion mass} the leading finite 
volume correction has already
been known for some time (Gasser and Leutwyler 1987):
\begin{equation}
M_\pi (L) = M_\pi \left\{ 1 +  \frac{M_\pi^2}{(4 \pi F_\pi)^2}
 \sum_{\vec{n} \neq \vec{0}} \, \frac{K_1(L |\vec{n}| M_\pi)}
              {L |\vec{n}| M_\pi} + O(M_\pi^4) \right\} \,.
\label{eq:mpigl}
\end{equation}
Moreover, neglecting pions which travel around the box more than once  
($|\vec{n}|>1$) L\"uscher (1986) has expressed the finite volume correction 
in terms of the $\pi \pi$ forward scattering amplitude $F(\nu)$, 
$\nu = (s-u)/(4 M_\pi)$:
\begin{equation}
M_\pi (L) - M_\pi = - \frac{3}{16 \pi^2 M_\pi L}
\int_{-\infty}^\infty \! \mathrm d y \, F(\mathrm i y) 
\mathrm e^{-\sqrt{M_\pi^2 + y^2}L} \,.
\label{eq:mpilu}
\end{equation}
The chiral expansion of $F(\nu)$ is known to $O(M_\pi^6)$ and
the corresponding coupling constants are reasonably well determined. 
Using this information one can work out the volume dependence of $M_\pi$
predicted by Eq.~(\ref{eq:mpilu}) (Colangelo and D\"urr 2004). The 
non-leading terms of the chiral expansion give non-negligible 
contributions, still the finite size effects are considerably 
underestimated by this approach, at least for $M_\pi > 500 \, \mbox{MeV}$. 
Probably terms with $|\vec{n}|>1$ are important for smaller volumes,
which is certainly the case in Eq.~(\ref{eq:mpigl}). However, 
Eq.~(\ref{eq:mpigl}) alone predicts even smaller finite size effects 
than Eq.~(\ref{eq:mpilu}). So it seems that one would need
higher orders in the chiral expansion as well as pions propagating
around the volume more than once, and the present understanding
of the finite size effects for $M_\pi$ is unsatisfactory.
For another investigation of hadron masses in a finite volume see
(Orth \etal 2004).

The volume dependence of $g_A$\index{axial charge} has recently 
attracted much interest. 
First calculations in chiral perturbation theory have been performed 
by Beane and Savage (2004). However, at the masses used in current 
simulations these leading-order formulae do not even reproduce the 
sign of the finite size effects observed in the Monte Carlo data, 
see, {\it e.g.}, (Ali~Khan \etal 2004b), (Sasaki \etal 2003). It remains to be
seen whether more advanced calculations in {\sc{cheft}} can solve this
discrepancy or whether lower quark masses are required.

\section{Conclusions}

What has been described here, are only the first steps of an ongoing
effort to combine calculations in {\sc{cheft}} and lattice simulations. The
overall impression is that the range of quark masses that can be used 
in actual Monte Carlo computations is beginning to overlap with the 
region of applicability of {\sc{chpt}}. In favourable
cases this overlap seems to be so large that meaningful fits are 
possible. Although fits without phenomenological input are still 
beyond reach, interesting information on (effective) coupling constants
can already be extracted. In most cases 
the leading chiral logarithm appears to be dominating only for 
rather small quark masses. Thus {\sc{cheft}} should be pushed to higher 
orders while on the lattice side results for lower quark masses are 
eagerly awaited.

\section*{Acknowledgements}

The studies reported in this paper have been performed within
the {\sc{qcdsf}}-{\sc{ukqcd}} collaboration. I wish to thank all my 
colleagues who have
contributed to this effort, in particular A.~Ali~Khan, Ph.~H\"agler, 
T.R.~Hemmert, R.~Horsley, A.C.~Irving, H.~Perlt, D.~Pleiter, 
P.E.L.~Rakow, A.~Sch\"afer, G.~Schierholz, A.~Schiller, H.~St\"uben 
and J.M.~Zanotti.

The numerical calculations have been performed on the Hitachi 
{\sc{sr8000}} at {\sc{lrz}} (Munich), on the Cray {\sc{t3e}} 
at {\sc{epcc}} (Edinburgh) and on the {\sc{ape}}mille at
{\sc{nic/desy}} (Zeuthen). This work has been supported in part by 
the {\sc{dfg}} (Forschergruppe Gitter-Hadronen-Ph\"anomenologie) 
and by the {\sc{eu}} Integrated
Infrastructure Initiative `Hadron Physics' as well as `Study of Strongly
Interactive Matter'.

\section*{References}

\frenchspacing
\begin{small}

\reference{Ali~Khan A \etal, 2002, \prd \vol 65 054505; \vol 67 059901 (E).}

\reference{Ali~Khan A \etal, 2004a, {\textit{Nucl Phys B}}
           \vol 689 175.} 

\reference{Ali~Khan A \etal, 2004b, arXiv:hep-lat/0409161.}

\reference{Allton C R \etal, 2002, \prd \vol 65 054502.}

\reference{Aoki S \etal, 2003, \prd \vol 68 054502.}

\reference{B\"ar O, 2004, arXiv:hep-ph/0409123.}

\reference{Beane S R and Savage M J, 2004, arXiv:hep-ph/0404131.}

\reference{Becher T and Leutwyler H, 1999, {\textit{Eur Phys J C}} 
           \vol 9 643.}

\reference{Colangelo G, 2004, arXiv:hep-ph/0409111.}

\reference{Colangelo G and D\"urr S, 2004, {\textit{Eur Phys J C}} 
           \vol 33 543.}

\reference{Colangelo G, Gasser J, and Leutwyler H, 2001, {\textit{Nucl Phys B}}
           \vol 603 125.} 

\reference{D\"urr S, 2003, {\textit{Eur Phys J C}} \vol 29 383.}

\reference{Gasser J and Leutwyler H, 1987, {\textit{Phys Lett B}}
           \vol 184 83.} 

\reference{Gasser J and Leutwyler H, 1988, {\textit{Nucl Phys B}}
           \vol 307 763.} 

\reference{G\"ockeler M \etal, 2004, arXiv:hep-ph/0409312.}

\reference{Hemmert T R, Procura M, and Weise W, 2003, \prd \vol 68 075009.}

\reference{Leinweber D B, Thomas A W, and Young R D, 2004, 
           \prl \vol 92 242002.}

\reference{Leutwyler H, 2000, arXiv:hep-ph/0008124.}

\reference{L\"uscher M, 1986, {\textit{Commun Math Phys}} \vol 104 177.}

\reference{Mei{\char"FF}ner U, 2000, arXiv:hep-ph/0007092.}

\reference{Orth B, Lippert T, and Schilling K, 2004, 
           {\textit{Nucl Phys Proc Suppl}} \vol 129 173.} 

\reference{Procura M, Hemmert T R, and Weise W, 2004, \prd \vol 69 034505.}

\reference{Sasaki S, Orginos K, Ohta S, and Blum T, 2003, \prd \vol 68 054509.}

\reference{Young R D, Leinweber D B, and Thomas A W, 2004,
           {\textit{Nucl Phys Proc Suppl}} \vol 128 227.}

\end{small}
\nonfrenchspacing

\end{document}